\documentclass[aps,prb,twocolumn,groupedaddress,showpacs]{revtex4}
\usepackage{hyperref}
\usepackage{amsmath}
\usepackage{graphicx}

\begin{document}

\title{The electronic structure of NaIrO$_3$, Mott insulator or band insulator?}
\author{Liang Du, Xianlei Sheng}
\affiliation{ Beijing National Laboratory for Condensed Matter Physics, 
              and Institute of Physics, 
              Chinese Academy of Sciences, 
              Beijing 100190, 
              China }
\author{Hongming Weng}
\email{hmweng@iphy.ac.cn}
\affiliation{ Beijing National Laboratory for Condensed Matter Physics, 
              and Institute of Physics, 
              Chinese Academy of Sciences, 
              Beijing 100190, 
              China }
\author{Xi Dai}
\email{daix@aphy.iphy.ac.cn}
\affiliation{ Beijing National Laboratory for Condensed Matter Physics, 
              and Institute of Physics, 
              Chinese Academy of Sciences, 
              Beijing 100190, 
              China }
\date{\today}

\begin{abstract}
     Motivated by the unveiled complexity of nonmagnetic insulating behavior in pentavalent post-perovskite NaIrO$_3$, we have studied 
     its electronic structure and phase diagram in the plane of 
     Coulomb repulsive interaction and spin-orbit coupling (SOC) by using the 
     newly developed local density approximation plus Gutzwiller method. Our theoretical study proposes the metal-insulator transition can be generated 
     by two different physical pictures: renormalized band insulator or Mott insulator regime. For the realistic material parameters in NaIrO$_3$, 
     Coulomb interaction $U=2.0 (J=U/4)$ eV and SOC strength $\eta=0.33$ eV, it tends to favor the renormalized band insulator picture as revealed
     by our study.
\end{abstract}


\maketitle

Recently there are increasing research activities on the spin-orbit 
coupling (SOC) driven metal-insulator transition (MIT) in 5d transition metal compounds.
\cite{Kim:prl2008,Kim06032009,Watanabe:prl2010,Pesin2010a,Lee:prb2012,Carter:arxiv:1207.2183}
Compared with 3d and 4d orbitals, the 5d ones are spatially more extended, leading to smaller on-site Coulomb repulsive interaction.
However, since the effect of Coulomb interaction is comparable with SOC, the interplay between these two factors can lead to
unexpected interesting phenomena in these materials.
\cite{Okamoto:prl2007,Shitade:prl2009,Yang:prb2010,Norman:prb2010,Podolsky:prb2011,Jiang:prb2011,WangFa:prl2011,Wan:prb2011,xu:prl2011,Arita:prl2012,subeti:prb2012}
The first well studied material of this type is Sr$_2$IrO$_4$,\cite{Kim:prl2008} 
where $t_{2g}$ and $e_g$ orbitals are separated by large 
crystal-field with the lower $t_{2g}$ orbitals filled by five electrons.
Due to strong SOC in the system, the $t_{2g}$ orbitals are split into nearly full-filled 
fourfold $j_{\text{eff}}=3/2$ states and half-filled twofold $j_{\text{eff}}=1/2$ states.
As a result, the system can be simplified to an effective one-band half-filled system with reduced bandwidth,
which has Mott insulator ground state with antiferromagnetic long range order.
Further theoretical studies show that both SOC and on-site Coulomb interaction are essential to
explain the Mott insulator behavior in this material, which has also been supported by recent experimental studies including
angle-resolved photoemission spectroscopy, optical conductivity, x-ray absorption\cite{Kim:prl2008} 
and resonant x-ray scatting\cite{Kim06032009} measurements.

Recently, another novel post-perovskite compound NaIrO$_3$ with pentavalent iridium Ir$^{5+}$ ions was 
synthesized by R. Cava's group in Princeton.\cite{Bremholm2011601}
The transport and susceptibility measurements show that NaIrO$_3$ is an insulator without magnetic order.
However, electronic structure calculations based on density functional theory (DFT) with local density approximation (LDA) 
found significant density of states (DOS) at the Fermi level, 
indicating the material to be metallic, which is in strong contrast with the experimental observations.  
The failure of DFT type calculation on predicting the basic electronic structure of the material implies
that the strong correlation effect may play an important role here, which is poorly treated by LDA alone.
The DFT calculation shows that in NaIrO$_3$ the energy bands near the Fermi level are mainly formed 
by the Ir $t_{2g}$ orbitals 
from its $5d$ shell. The SOC splits the $t_{2g}$ bands into 
two groups of bands with effective total angular momentum 
$j_{\text{eff}}=1/2$ and $j_{\text{eff}}=3/2$ respectively. 
Since for NaIrO$_3$ there are totally four electrons remaining in the $t_{2g}$ bands, 
large enough SOC will naturally lead to band insulator phase with fully occupied $j_{\text{eff}}=3/2$ bands and 
empty $j_{\text{eff}}=1/2$ bands. 
While the LDA calculation predicts a metallic phase, simply because the spin-orbit splitting is still several times 
smaller than that of the $t_{2g}$ bandwidth and is not strong enough to generate a band insulator by SOC alone. 

In this type of system, there are two possible ways for the correlation effect induced by local Coulomb interaction 
to generate insulating behavior. The first one is called ``renormalized band insulator", in which 
the correlation effect reduces the effective bandwidth and enhances the effect of SOC leading to 
a band insulator phase with ``renormalized" band 
structure. The second possibility is the Mott insulator with completely vanishing of quasiparticles, 
which is caused by strong local Coulomb repulsive interaction.
Unlike the Mott insulator phase in Sr$_2$IrO$_4$, which has five $5d$ electrons on each Ir ion leading to
magnetically ordered phase at zero temperature, in NaIrO$_3$ there are four electrons in the $t_{2g}$ orbitals,
which leads to nonmagnetic atomic ground states with spin and orbital moments canceling each other 
even in the strong coupling limit with very large Coulomb interaction and weak SOC.  Since there is no symmetry 
difference between the above two possible insulator phases, these two ``phases" should be adiabatically connected 
to each other and there will be no phase transition but crossover between them in the parameter space. While 
it is then interesting to ask that for this particular material NaIrO$_3$, is it more close to a 
``renormalized band insulator" or ``Mott insulator"?
 
In the present letter, we apply the newly developed LDA+Gutzwiller
\cite{bunemann:prb1998, Lechermann:prb2007, Bunemann:prl2008,Deng:epl2008,wang:prl2008, Deng:PhysRevB2009,tian:prb2011,Lanata:prb2012} method to study the interplay between SOC and 
local Coulomb interaction among $5d$ electrons in NaIrO$_3$. We find that both the above two effects are important to 
explain the insulating behavior in this material and this particular material would be better described by the 
picture of ``renormalized band insulators" than that of Mott insulator.


\begin{figure}[ht]
\centering
\includegraphics[scale=0.31]{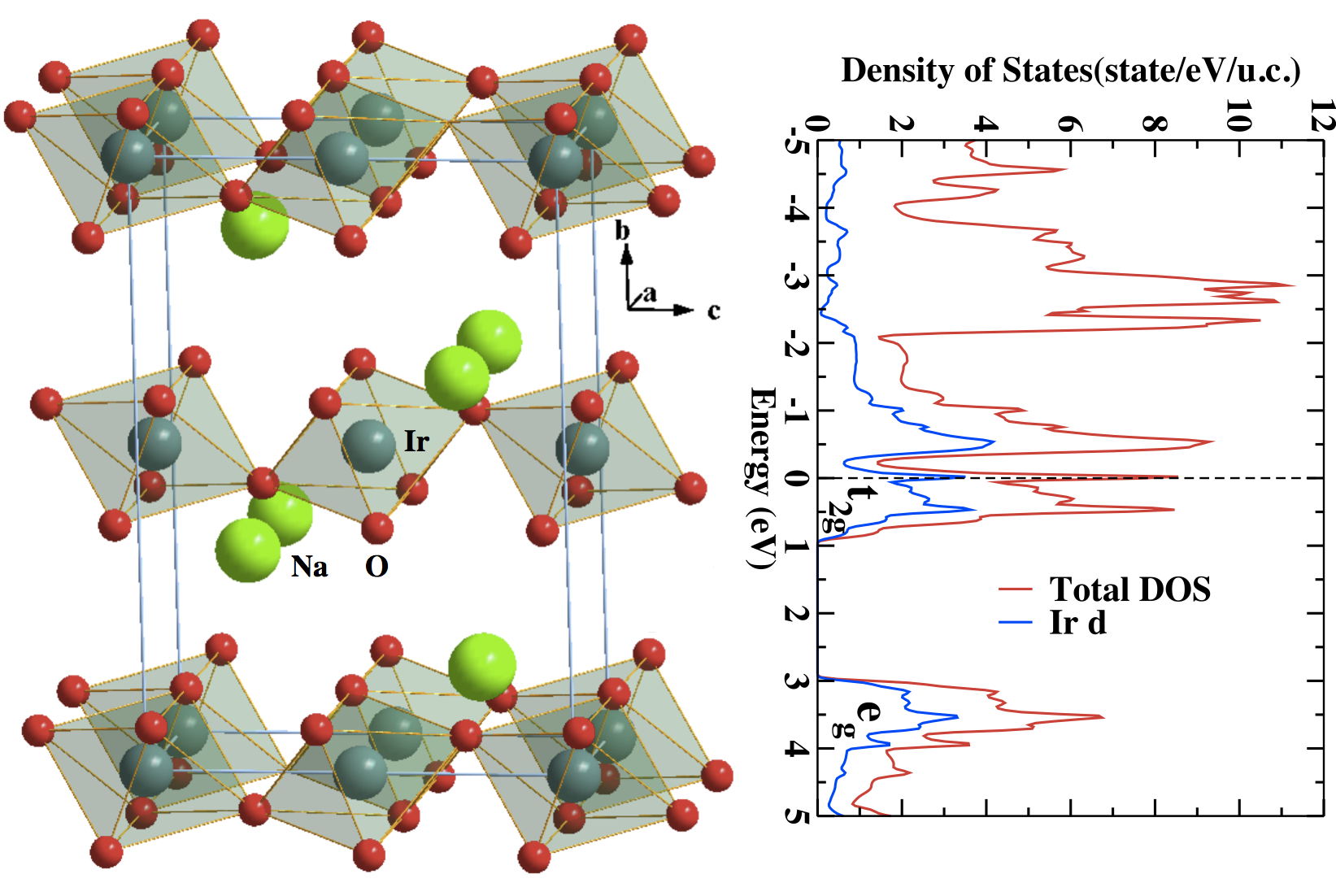}
\caption{(Color online) 
The experimental crystal structure of NaIrO$_3$ is shown in the left panel. The green (large) ball represents Na atom. Red (small) and darkslategray (middle) balls represent O and Ir atoms, respectively. Total density of states and the projected one for Ir $d$ orbitals obtained from LDA calculation of 
NaIrO$_3$ are shown in the right panel.
\label{struct}}
\end{figure}


%
%
Firstly, we carry out the electronic structure calculation by DFT with LDA using the experimentally determined post-perovskite 
crystal structure,\cite{Bremholm2011601} 
which is shown in Fig.\ref{struct}. In such structure each Ir atom is surrounded by six oxygen to form an octahedron. 
These octahedrons are connected by sharing conner oxygen along $c$-axis and by sharing edge oxygens along $a$-axis, which forms 
IrO$_3$ sheet stacking along $b$-axis separated by Na ion layer. Our LDA calculation indicates that the coupling between neighboring 
IrO$_3$ sheets are quite weak, being consistent with the two-dimensional electron transporting property found in transport experiments. 
To search for the possible magnetic ordering of Ir ions, four types of magnetic orderings, namely, ferromagnetic 
ordering in both $a$- and $c$-axis (noted as aFcF), ferromagnetic ordering along $a$- while anti-ferromagnetic ordering along $c$-axis (noted as aFcA), 
anti-ferromagnetic ordering along $a$- and ferromagnetic along $c$-axis (noted as aAcF), and anti-ferromagnetic ordering along both $a$- and $c$-axis (noted as aAcA), 
are studied. The first-principles calculations of these states are performed by using full potential all electron method implemented in WIEN2k 
software package in order to get highly accurate total energies. The LDA calculations can get only nonmagnetic solution and density of states 
in Fig.\ref{struct} show that Ir 5d $t_{2g}$ and $e_g$ orbitals are well separated by about 3.5 eV under octahedral crystal-field. 

The
bandwidth of $t_{2g}$ orbitals are about 3.0 eV, which is quite large since the direct overlap of neighboring $t_{2g}$ orbitals is possible due to the 
edge-sharing connection along $a$-axis. The main physics is dominated by Ir $t_{2g}$ orbitals around the Fermi level. 
Since both the correlation effect and the SOC are important in $5d$ transition metal compounds, we further apply the LDA+U+SOC to study the electronic
structure of this material with U varying from $1.0$ to $7.0$ eV. Our numerical results show that 
the LDA+U+SOC calculations always converge to nonmagnetic and metallic solution even when U is as large as 7.0 eV. 
These results indicate the failure of mean field treatment of such a correlated system and profound influence of SOC on the electronic structures.
In order to well describe the observed insulating behavior in NaIrO$_3$, we have further performed LDA+Gutzwiller calculations combining the 
DFT with the Gutzwiller variational method, which can treat the correlation effects more precisely.


We firstly construct an accurate low energy model
Hamiltonian, which can well catch the crystal-field splitting,
the hopping parameters among active $t_{2g}$ local orbitals and the SOC in
real material. To do this, the projected atomic Wannier
functions\cite{mlwf:prb1997, mlwf:prb2001, weng:prb2009} (PAW) are constructed for the $t_{2g}$ orbitals of
Ir ion by using OpenMX software package\cite{openMX} within LDA
calculation. This approach has been used to treat 3d
transition metal $t_{2g}$ orbitals successfully.\cite{weng:prb2010} In one unit
cell, there are two non-equivalent Ir ions and the $t_{2g}$ orbitals are defined in each local coordinates. The obtained
PAW orbitals are quite localized and the band structure
obtained by first-principles calculations can be well reproduced by
the tight-binding Hamiltonian using these basis. It is shown that the
$d_{xy}$-like PAW is lower than the nearly degenerated $d_{xz}$- and $d_{yz}$-like PAWs by about 0.52 eV, which is 
known as tetragonal crystal-field. The atomic SOC
is added to the above tight-binding (TB) Hamiltonian and the effective SOC
strength $\eta$ is found to be 0.33 eV when fitted with the
LDA+SOC calculations. As a result, 
the effect of SOC splits $j_{\text{eff}}=1/2$ (higher in energy) and $j_{\text{eff}}=3/2$ states by about $3\eta/2 \simeq 0.5$ eV. 
Therefore, the strength of SOC and tetragonal crystal-field are comparable and they compete against each other in 
considering of the orbital degeneracy.
Implemented with the local Coulomb interaction terms among the $t_{2g}$ orbitals, the total Hamiltonian can be written as,
\begin{align}
   H = H_{t} + H_{u} + H_{\eta} + H_{\Delta}.
\end{align}
The first term describes the hopping process of electrons between local spin-orbitals ``$a\sigma$" and ``$b\sigma'$",
\begin{equation}
     H_{t} =  \sum_{i \neq j}\sum_{a\sigma, b\sigma'}t_{ia,jb}^{\sigma\sigma'} d_{i,a\sigma}^{\dagger}d_{j,b\sigma'}
\end{equation}
where $\sigma$ denotes electronic spin, and $a$ represents the three $t_{2g}$ orbitals with 
$a=1,2,3$ corresponding to $d_{yz}, d_{zx}, d_{xy}$ orbitals respectively.
The rest terms of the Hamiltonian are all local terms expressed by $H_{loc}^{i}=H_{u}^{i}+H_{\eta}^{i} + H_{\Delta}^{i}$,
which contains Coulomb interaction $H_u^i$, SOC $H_{\eta}^i$ and tetragonal crystal-field 
splitting $H_{\Delta}^{i}$. 
(In the following, the site index is suppressed for sake of simplicity).
\begin{eqnarray}
 H_u & = &U \sum_{a} n_{a\uparrow} n_{a\downarrow}
       + U' \sum_{a<b,\sigma\sigma'} n_{a\sigma} n_{b\sigma'}
       - J  \sum_{a<b,\sigma} n_{a\sigma} n_{b\sigma} \nonumber\\
     && - J \sum_{a<b} \left(
         d_{a\uparrow}^\dagger d_{a\downarrow} d_{b\downarrow}^\dagger d_{b\uparrow}
       + d_{a\uparrow}^\dagger d_{a\downarrow}^\dagger d_{b\uparrow} d_{b\downarrow} + h.c. \right),
\end{eqnarray}
\begin{equation}
 H_{\eta} = \sum_{a\sigma, b\sigma'} \eta \langle a\sigma | l_x s_x + l_y s_y 
         + l_z s_z |b\sigma' \rangle d_{a\sigma}^\dagger d_{b\sigma'},
\end{equation}
\begin{equation}
 H_{\Delta} = \sum_{a\sigma, b\sigma'} \Delta_{a\sigma, b\sigma'} d_{a\sigma}^{\dagger} d_{b\sigma'},
\end{equation}
where  $U$ ($U'$) is the strength of intra-orbital (interorbital) Coulomb interaction and  $J$ describes the Hund's rule coupling.
$U$, $U'$ and $J$ satisfy Kanamori constraint $U=U'+2J$. 
$l$ and $s$ represent the orbital and spin angular momentum operators, respectively.
$\eta$ represents the SOC strength and $\Delta$ describes the crystal-field splitting between the three $t_{2g}$ orbitals.
We emphasize that in the present study we treat the Hund's rule coupling terms in a full rotational invariant way, which
is crucial to guarantee that the locking of spin and orbital spaces is purely due to SOC but not the $S_\text{z}$-$S_\text{z}$ Hund's coupling term
with artificially chosen z-direction. We also fix the ratio between Hund's coupling $J$ and Hubbard interaction $U$ to be $1/4$ throughout
the entire paper.

Next, we briefly introduce the Gutzwiller wave function (GWF) used in this paper. The generalized GWF
$|\Psi_{\text{G}}\rangle$ with rotational invariant local interaction terms can be constructed
by acting a many-particle projection operator $\mathcal{P}$
on the uncorrelated wave function $|\Psi_{0}\rangle$, 
\begin{equation}
    |\Psi_{\text{G}}\rangle=\mathcal{P}|\Psi_{0}\rangle,
\end{equation}
with  
\begin{equation}
    \mathcal{P} = \prod_{\bf{R}}\mathcal{P}_{\bf{R}}=\prod_{\bf{R}}\sum_{\Gamma\Gamma'}
          \lambda({\bf{R}})_{\Gamma\Gamma'}|\Gamma, {\bf{R}}\rangle\langle\Gamma',{\bf{R}}|.
\end{equation}
where $|\Psi_{0}\rangle$ is a normalized uncorrelated wave function in which
Wick's theorem holds, $|\Gamma,\bf{R}\rangle$ represents atomic eigenstates on
site $\bf{R}$ and $\lambda({\bf{R}})_{\Gamma\Gamma'}$ are Gutzwiller variational parameters to be determined
by variational principle.
In our work, $|\Gamma, \bf{R}\rangle$ are eigenstates of atomic Hamiltonian $H_{loc}$. 
The expectational value of hopping terms $H_t$ in our Hamiltonian can be expressed as:
\begin{equation}
     \langle\Psi_G|H_t|\Psi_G\rangle = \sum_{ij}\sum_{\alpha\beta}\sum_{\delta\gamma} 
     t_{ij}^{\alpha\beta} \mathcal{R}_{\alpha\gamma}^\dagger \mathcal{R}_{\delta\beta} 
     \langle\Psi_0|d_{i\gamma}^\dagger d_{j\delta} |\Psi_0\rangle
\end{equation}
with $\alpha (\beta,\delta,\gamma)$ being combined spin-orbital index and 
\begin{equation}
\mathcal{R}_{\alpha\gamma}^{\dagger}
=\frac{\text{\text{Tr}\ensuremath{\left(\phi^{\dagger}{d}_{\alpha}^{\dagger}\phi{d}_{\gamma}\right)}}}
      {\sqrt{n_{\gamma}^{0}(1-n_{\gamma}^{0})}}, 
\end{equation}
\begin{equation}
\phi_{II'}=\langle I|\mathcal{P}|I'\rangle\sqrt{\langle\Psi_{0}|I'\rangle\langle I'|\Psi_{0}\rangle},
\end{equation}
where $|I\rangle$ stand for the many-body Fock states and $n_{\gamma}^0 = \langle \Psi_0| n_{\gamma}|\Psi_0 \rangle$ with $n_{\gamma}$ being
occupation number operator for ${\gamma}$ state.
The Gutzwiller variational parameters $\lambda({\bf{R}})_{\Gamma\Gamma'}$ are determined by minimizing the above total energy.
The quasiparticle weights in the generalized Gutzwiller method is defined as the eigenvalues of the Hermite 
matrix $\mathcal{R}^{\dagger}\mathcal{R}$. The detail numerical procedure for the rotational invariant Gutzwiller method can be
found in reference 
[\onlinecite{bunemann:prb1998,Lechermann:prb2007,Bunemann:prl2008,Deng:epl2008,wang:prl2008, Deng:PhysRevB2009,tian:prb2011,Lanata:prb2012}].
%
\begin{figure}[ht]
\centering
\includegraphics[scale=0.30]{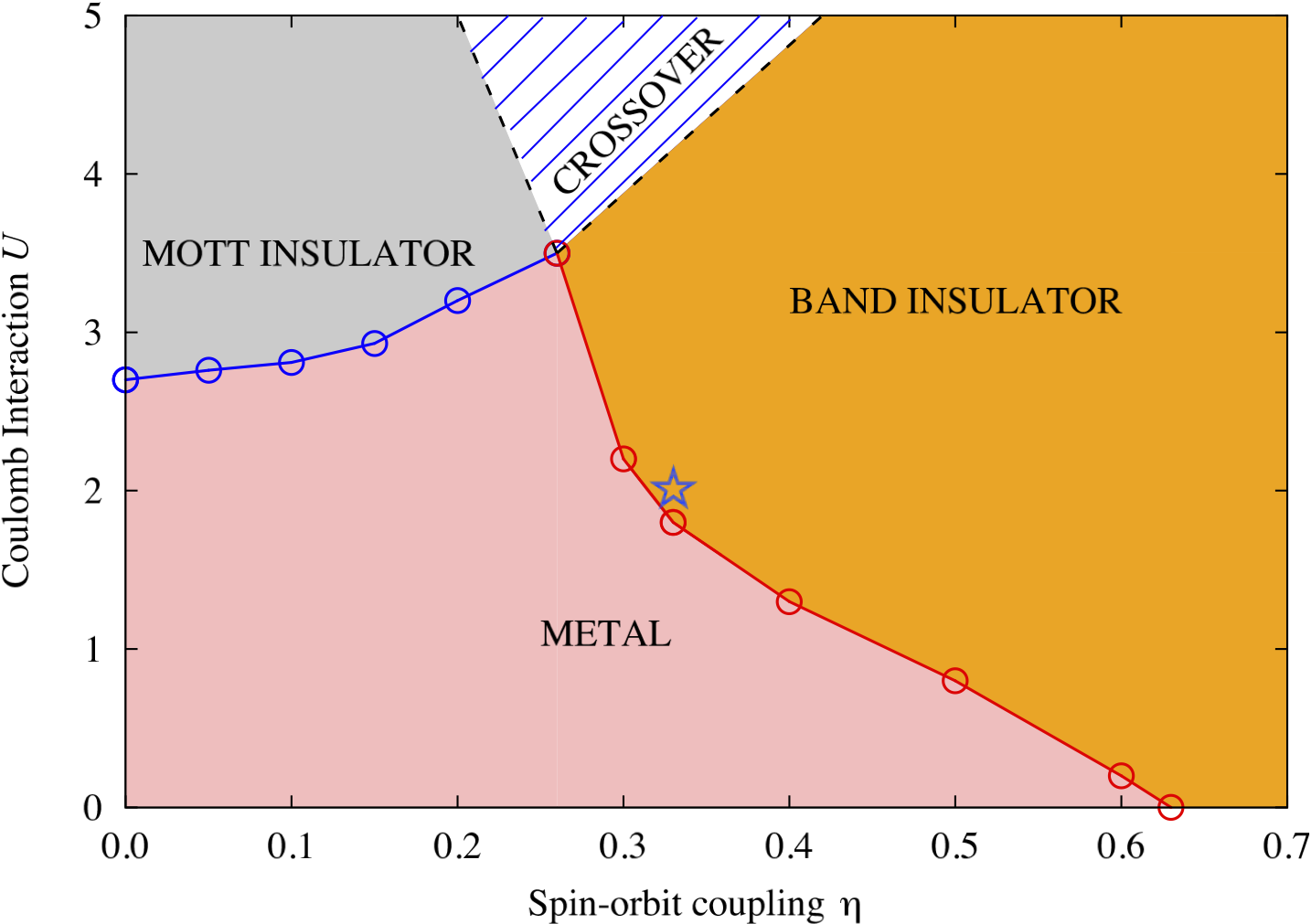}
\caption{(Color online) The phase diagram in the plane of Coulomb interaction and spin-orbit coupling (SOC). There exists three different 
regions: Mott insulator, Band insulator and Metal. The blue star locating in band insulating region denotes the realistic 
parameters of NaIrO$_3$ with Coulomb repulsive interaction $U=2.0$ eV and SOC strength $\eta=0.33$ eV.
The Hund's rule coupling $J$ is fixed as $U/4$.
\label{phase-diagram}}
\end{figure}
\begin{figure}[ht]
\centering
\includegraphics[scale=0.30]{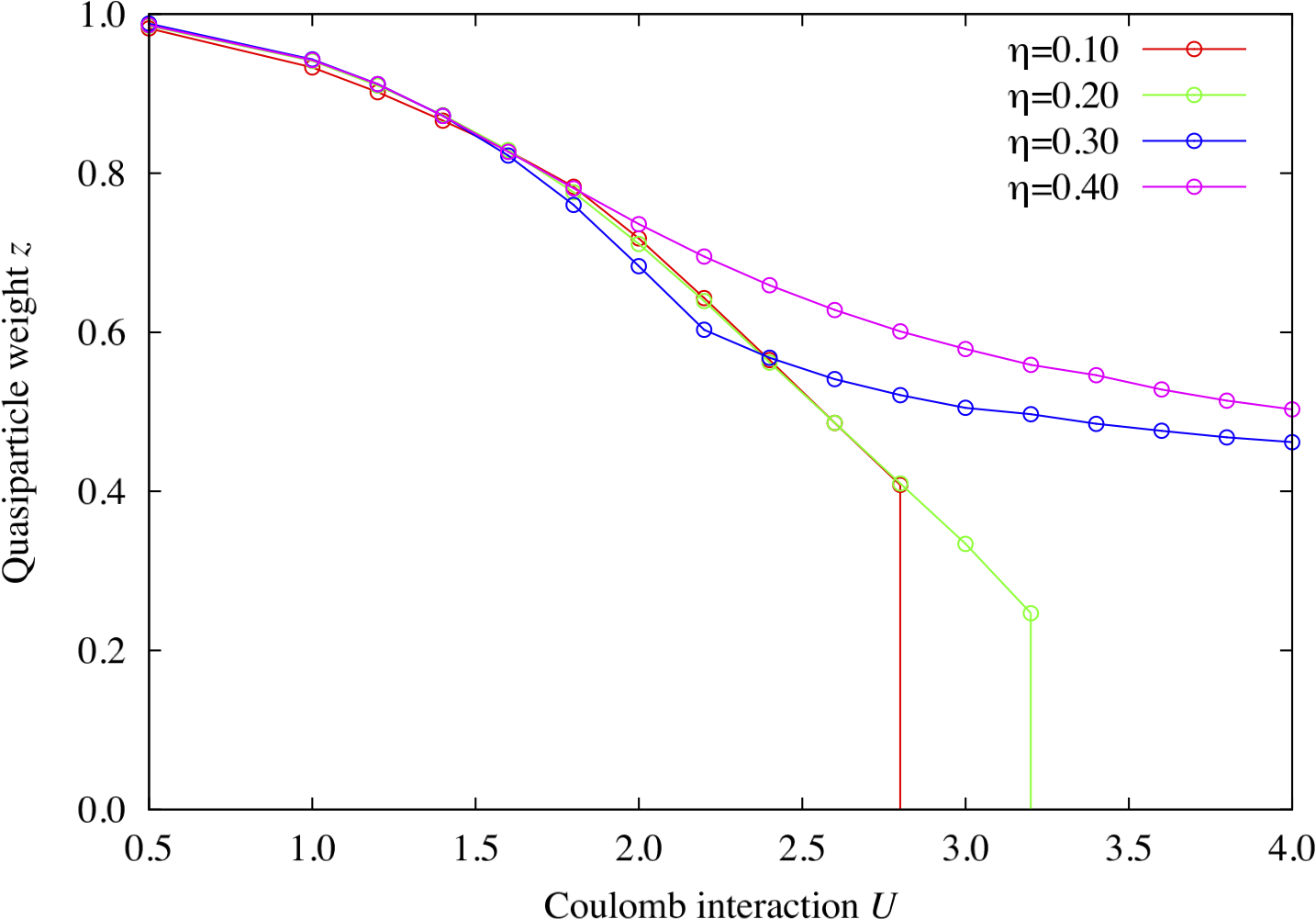}
\caption{(Color online) The quasiparticle weight $z$ as a function Coulomb interaction for different SOC strength $\eta$. 
Note here the orbital is selected to be the one with smallest quasiparticle weight.
For $\eta=0.1 (0.2)$ eV, there exists a first order transition to Mott insulator around $U=2.8 (3.2)$ eV. While for $\eta=0.3(0.4)$ eV, 
the transition to band insulating state is around $U=2.2 (1.2)$ eV.
The Hund's rule coupling $J$ is fixed as $U/4$.
\label{zvec}}
\end{figure}
\begin{figure}[ht]
\centering
\includegraphics[scale=0.68,angle=0]{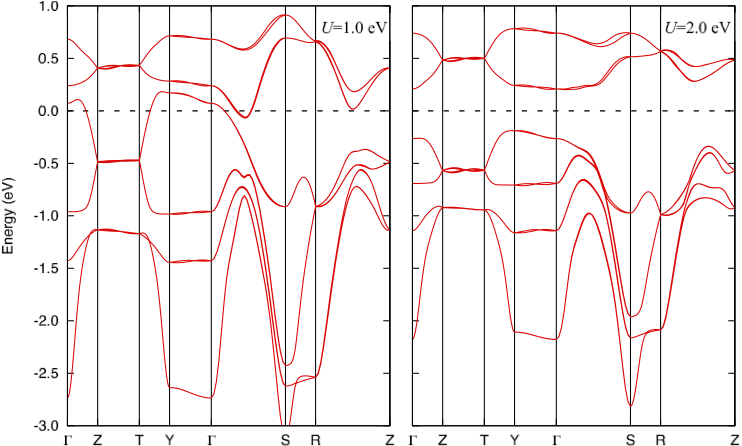}
\caption{(Color online) At realistic SOC strength $\eta=0.33$ eV, the band structure of NaIrO$_3$ with $U$=1.0 eV (left panel) and 2.0 eV (right panel) 
are calculated. For $U=2.0$ eV case, clearly there exists a gap of about $500$ meV. 
The Hund's rule coupling $J$ is fixed as $U/4$.
\label{band-structure}}
\end{figure}
The complete phase diagram in the parameter space spanned by SOC $\eta$ and Hubbard repulsive interaction U with fixed $J/U=1/4$ has been 
plotted in Figure.\ref{phase-diagram}. According to the strength of SOC, the whole phase diagram can be divided into two 
regions. In the left part of the phase diagram, the SOC is much weaker than the tetragonal crystal-field, which splits the $t_{2g}$ 
orbitals into two-fold (including spin) low lying and four-fold high lying energy levels, respectively. 
With the lower two-fold bands being fully occupied, the rest of the 
system can be considered effectively as two-band system filled by two electrons. A typical Mott transition happens when 
the Hubbard repulsive interaction U increases over the critical value, which is around $2.7$ eV in the weak SOC limit. 
The quasiparticle weight $z$ as a function of $U$ for the weak SOC strength ($\eta=0.1, 0.2$ eV) is plotted in Fig.\ref{zvec}, which shows 
typical Mott transitions with strong first order nature.  The behavior of the quasiparticle weight, both the concave 
shape of the curve and the first order nature for the Mott transition, agrees quite well with results 
obtained by studying the model Hamiltonians with semi-circle like density of states. The increment of SOC will further split the 
above two bands and transfer part of the electrons from one band to another, which will suppress the orbital correlation in the 
effective two-band system and raise the critical $U_c$ for the system. The same increment of $U_c$ as the function of energy 
level splitting in the two-band Hubbard model has been also reported and discussed in detail by P. Werner et al.\cite{Werner:prl2007}
 
On the right part of the phase diagram, compared to the  tetragonal crystal-field, the SOC becomes more dominant and
we find a completely deferent behavior as we increase the interaction strength $U$. The quasiparticle weight $z$ in the large SOC region 
($\eta=0.3,0.4$ eV) is again
plotted as the function of $U$ in Fig.\ref{zvec}, which indicates that the quasiparticles with non-zero weight $z$ persists all the way up to
above $4.0$ eV. Detailed analysis of the corresponding band structure, which is plotted in Fig.\ref{band-structure} for SOC strength $\eta=0.33$ eV,
indicates that there is a transition to band insulator around $U_c=1.8$ eV, above which a clear band gap appears between the forth and fifth
bands as illustrated in Fig.\ref{band-structure}. Unlike the Mott transition we discussed in the previous paragraph, 
the transition in the right part of 
our phase diagram is a typical Lifshitz transition characterized by the vanishing of 
the electron and hole fermi surfaces right at the transition point. 
Therefore for this particular material, with the different strength of SOC, the on-site Coulomb interaction can induce Mott transition for
weak SOC and Lifshitz transition for strong SOC cases. The realistic SOC strength of NaIrO$_3$ has been fitted to be around $0.33$ eV, which
is indicated in Fig.\ref{phase-diagram} by a blue star and clearly located at the right side of the phase diagram.

In conclusion, our LDA+Gutzwiller calculation confirms that NaIrO$_3$ is a band insulator with renormalized bandwidth induced by the correlation
effect among the $5d$ orbitals. As shown in Fig.\ref{zvec}, the quasiparticle weight drops to about $0.65$ at $U=2.0$ eV, which completely delimitates
the overlap between the conduction and valence band leading to a band insulator with renormalized bandwidth. The SOC in this material plays
a very important role, which leads to very different behavior for weak and strong SOC. In the former case, the on-site Coulomb interaction among
the $5d$ orbitals will generate a Mott transition, after which the system is well described by local moments. While in the latter case, when the strength
of SOC overwhelms the tetragonal crystal-field, the correlation effect will always generate band insulators with renormalized bandwidth. The correlation
effect in these renormalized band insulator phase may also manifest itself in the dynamical properties, i.e. the unusual exciton behavior, which will be
discussed in our further publications.

This work was supported by the National Science Foundation of China and from the 973 program of China under 
Contract No. 2011CBA00108. We acknowledge the helpful discussion with professor Y.B. Kim and G. Cao.

\bibliography{NaIrO3}
\end{document}